\documentclass{elsart}

\usepackage{amsmath}
\usepackage{graphicx}

\begin{document}

\begin{frontmatter}

\title{Phase diagrams of half-filled 1D and 2D extended Hubbard model within COM}

\author{Adolfo Avella and Ferdinando Mancini}

\address{Dipartimento di Fisica ``E.R. Caianiello'' - Universit\`{a} degli
Studi di Salerno \\ Unit\`{a} di ricerca INFM di Salerno -
Laboratorio Regionale INFM ``SUPERMAT''
\\ Via S. Allende, I-84081 Baronissi (SA), Italy}

\begin{abstract}
The half-filled extended Hubbard model, in one and two dimensions,
is studied by means of the 2-pole approximation within the
Composite Operator Method with the aim at improving the
possibilities to describe some of the experimental features
observed for quasi-1D organic superconductors and Cu-O planes of
cuprates. The phase diagrams ($T$-$V$ and $V$-$U$) are analyzed
with respect to the paramagnetic metal - paramagnetic insulator -
charge ordered phase transitions. The relevant features of the
diagrams (rank of the phase transitions, critical points,
reentrant behavior) are discussed in detail.
\end{abstract}

\end{frontmatter}

\section{Introduction}
\label{Intro}

The extended Hubbard model \cite{Emery:79} is the simplest
Hamiltonian that takes into account non-local Coulomb
interactions. Its relevance to the description of cuprate
superconductors, fullerides and materials presenting charge
ordered phases (rare-earth pnictides, vanadium oxides, manganites,
magnetite, Bechgaard salts, \ldots) is widely recognized in the
literature \cite{Varma:87,Janner:95,McKenzie:01,Calandra:02}.

In this paper, we study the model, at half-filling in one and two
dimensions, by means of the 2-pole approximation within the
Composite Operator Method (COM) (see Ref.~\cite{Mancini:00} and
references therein). The latter, starting from the consideration
that the original electrons lose their identity in presence of
strong correlations, uses composite operators (i.e., operators
built as products of the electronic ones) to describe the
effective elementary excitations of the interacting system under
analysis. The non-canonical algebra satisfied by the composite
operators dictates constraints (i.e., Algebra Constraints
\cite{Mancini:00}) that are systematically used to fix the
representation where the propagators are realized and
self-consistently compute the unknown parameters appearing in the
theory \cite{Mancini:00}.

In the next sections, we will, first, briefly present the model
and the method and, then, discuss the phase diagrams obtained by
the application of the latter to the former. The rank of the
transitions between homogeneous and charge ordered phases are
studied together with the relation between such transitions and
the paramagnetic metal-insulator one. The relevant features of the
diagrams (critical points and reentrant behavior) are also
analyzed in detail.

\section{Hamiltonian, field equations and solution}

The extended Hubbard model reads as
\begin{equation}
H=\sum_{\bf i} [-\mu c^\dagger (i)c(i)-2dtc^\dagger (i)c^\alpha
(i)+ Un_\uparrow (i)n_\downarrow (i)+dVn(i)n^\alpha (i)]
\end{equation}
where $\mu$ is the chemical potential,
$c^\dagger(i)=(c_\uparrow^\dagger(i) \, c_\downarrow^\dagger(i))$
is the creation electronic operator in spinorial notation,
$i=(\mathbf{i}, t)$, $\mathbf{i}$ is a lattice vector of a
$d$-dimensional square lattice, $t$ is the hopping integral, $U$
is the onsite Coulomb interaction, $V$ is the intersite one,
$n(i)=n_\uparrow(i)+n_\downarrow(i)$, $n_\sigma(i)$ is the number
operator for electrons of spin $\sigma$. Hereafter, the hopping
integral will be used as reference unit for all energies. We have
used the notation $\phi^\alpha (\mathbf{i}, t)= \sum_{\bf j}
\alpha_\mathbf{ij} \phi (\mathbf{j}, t)$, where $\phi$ can be any
operator and $\alpha_\mathbf{ij}$ is the projector on the first
$2d$ neighbor sites on the lattice. We have
$\alpha(\mathbf{k})=\mathcal{F}[\alpha_\mathbf{ij}]=1/d\sum_{n=1}^d
\cos(k_n)$, where $\mathcal{F}$ is the Fourier transform.

We have chosen, as basic field, $\psi^\dagger(i)= (\xi^\dagger (i)
\, \eta^\dagger (i))$, where $\xi(i)=n(i)c(i)$ and
$\eta(i)=c(i)-\xi(i)=[1-n(i)]c(i)$ are the Hubbard operators. They
satisfy the following equations of motion
\begin{equation}
\mathrm{i}{\partial \over {\partial t}}\psi (i) = \begin{pmatrix}
{-\mu \xi (i)-2d[tc^\alpha (i)+t\pi (i)-V\xi (i)n^\alpha (i)]}\\
{-(\mu -U)\eta (i)+2d[t\pi (i)+V\eta (i)n^\alpha (i)]}
\end{pmatrix}
\end{equation}
where $\pi (i)={1 \over 2}\sigma ^\mu n_\mu (i)c^\alpha
(i)+c(i)c^{\alpha^\dagger }(i)c(i)$, $\sigma
^\mu=(-1,\vec{\sigma})$, $n_\mu(i)=(n(i),\vec{n}(i))$ is the
charge and spin number operator, $\vec{n}(i)=c^\dagger(i)
\vec{\sigma} c(i)$ and $\vec{\sigma}$ are the Pauli matrices.

After the choice we made for the basis, we have computed, in the
2-pole approximation within COM \cite{Mancini:00}, the retarded
thermal Green's function $G({\bf k},\omega)=\mathcal{F}\langle
\mathcal{R}[\psi (i)\psi ^\dagger (j)] \rangle$ ($\mathcal{R}$ is
the retarded operator)
\begin{equation}\label{2pole}
G(\omega, {\bf k})=\sum_{i=1}^2\frac{\sigma^{(i)}({\bf k})}{\omega
-E_{i}({\bf k})+\mathrm{i}\delta}.
\end{equation}
where $E_{i}({\bf k})$ are the eigenvalues of the energy matrix
$\epsilon({\bf k})=m({\bf k})I^{-1}({\bf k})$, $I({\bf
k})=\mathcal{F}\langle \{ \psi({\bf i},t), \psi^{\dagger}({\bf
j},t) \} \rangle$ is the normalization matrix of the basis and
$m({\bf k})=\mathcal{F}\langle \{
\mathrm{i}\frac{\partial}{\partial t}\psi({\bf i},t),
\psi^{\dagger}({\bf j},t)\} \rangle$. The spectral weights
$\sigma^{(i)}({\bf k})$ can be computed as
\begin{equation} \label{Eq.B.19}
\sigma^{(i)}_{ab}({\bf k})=\sum_{c=1}^n \Omega_{ai}({\bf
k})\Omega_{ic}^{-1}({\bf k})I_{cb}({\bf k}) \quad a,b=1,\ldots,n
\end{equation}
where the matrix $\Omega ({\bf k})$ has the eigenvectors of
$\epsilon({\bf k})$ as columns \cite{Mancini:00}. For the sake of
brevity, we do not report the complete expressions of $E_i$ and
$\sigma^{(i)}$, which can be found in Ref.~\cite{Avella:04}. The
parameters appearing in the theory, and not connected to the
Green's function under analysis ($\langle n(i)n^\alpha (i)
\rangle$ and ${1 \over 4} \langle n_\mu ^\alpha (i)n_\mu
(i)\rangle - \langle [c_\uparrow (i)c_\downarrow (i)]^\alpha
c_\downarrow ^\dagger (i)c_\uparrow ^\dagger (i) \rangle$), will
be self-consistently computed (the first) by calculating the
density-density correlation function within the one-loop
approximation \cite{Mancini:95b} and (the second) by means of the
local algebra constraint \cite{Mancini:00} $\langle \xi(i)
\eta^\dagger(i) \rangle=0$.

The set of self-consistent equations, fixing the parameters
appearing in the theory, is highly non-linear. According to this,
it is not strange at all that it admits two distinct solutions
that, hereafter, we will call COM1 and COM2. The Composite
Operator Method tries to give answers in the whole space of model
and physical parameters and the presence of two solutions should
be seen as a richness of the method. For the one-dimensional
system we will consider only COM2 solution as with this, for the
simple Hubbard model, we got excellent agreements with the Bethe
Ansatz exact solution \cite{Avella:00}. For the two-dimensional
case, we will study both solutions as they will permit us to
analyze two different behaviors that could be both observed
experimentally.

\section{Phase diagrams}

\subsection{One-dimensional system}

\begin{figure}[tbp]
\begin{center}
\includegraphics[width=6.5cm,clip=]{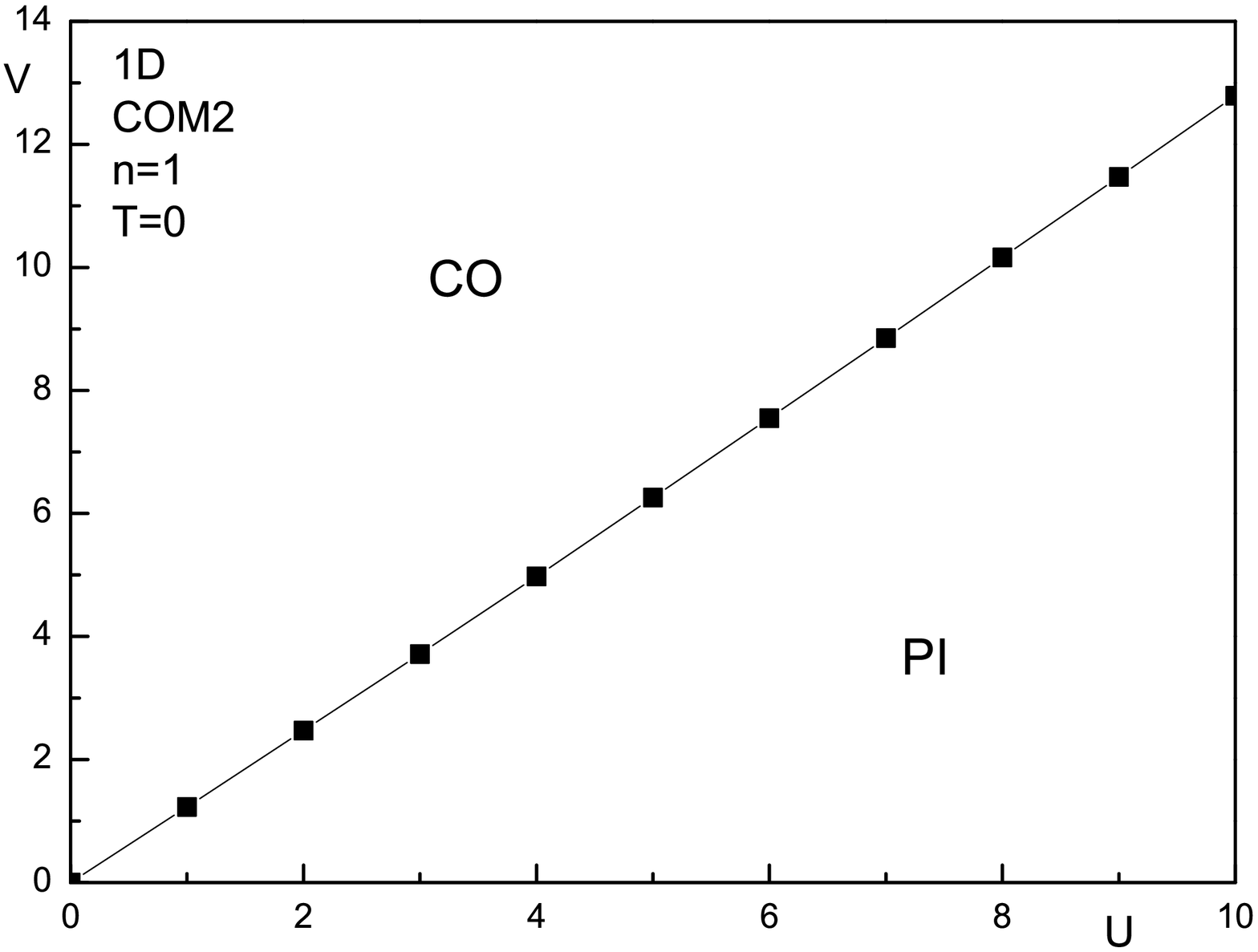}
\hfill
\includegraphics[width=6.5cm,clip=]{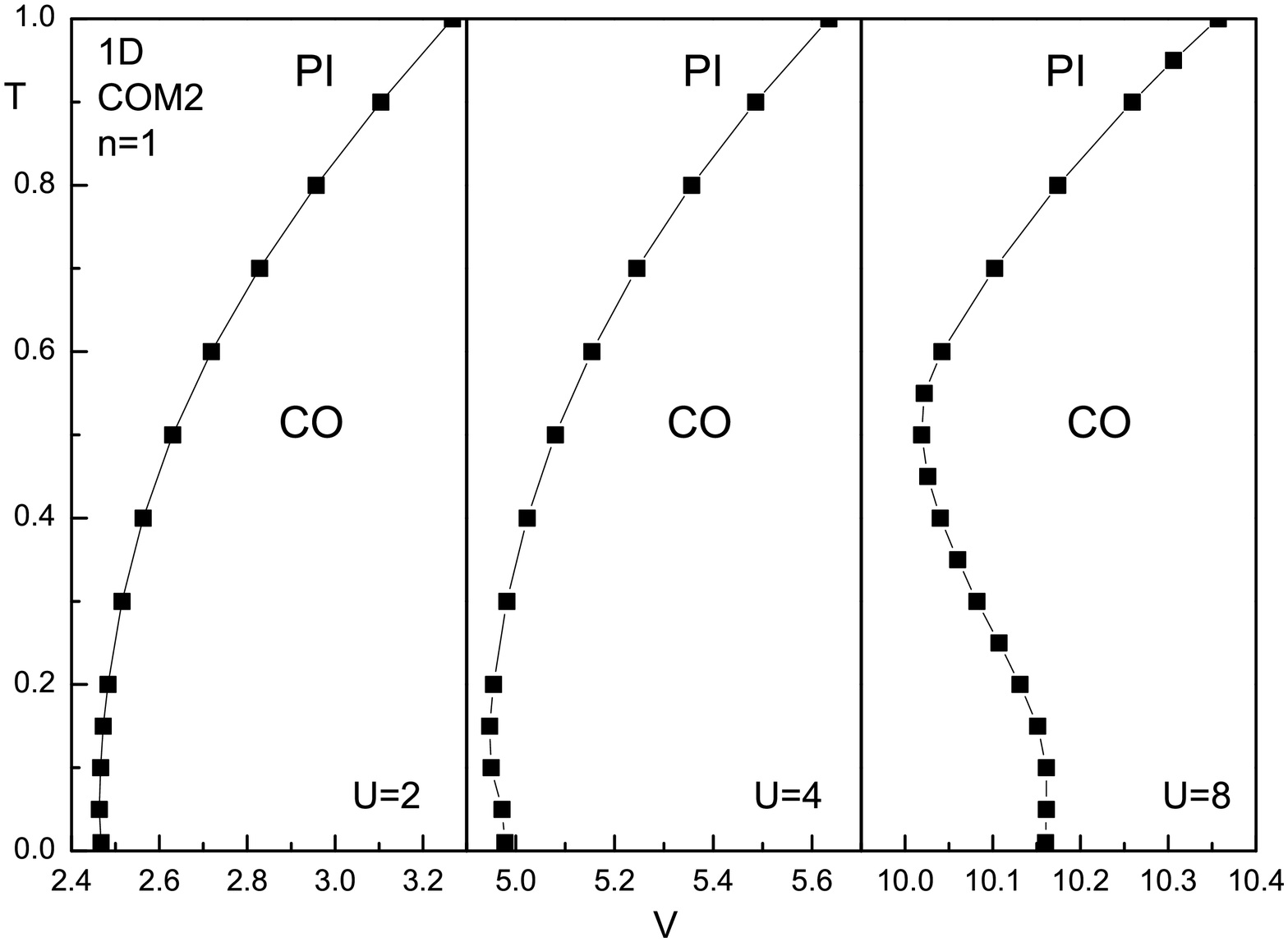}
\end{center}
\caption{(left) The phase diagram $V$-$U$ at zero temperature (PI
= paramagnetic insulator, CO = charge ordered phase). (right) The
phase diagram $T$-$V$  at $U=2$, $4$ and $8$.} \label{Fig1-2}
\end{figure}

COM2 solution for the one dimensional system reports a phase
transition from the Mott insulating phase to a inhomogeneous
charge ordered state of checkerboard type for positive values of
the intersite Coulomb potential greater than some critical one.
This result is consistent with many other studies
\cite{Hirsch:84a,Nakamura:00,Jeckelmann:02}. The nature of this
phase transition is not well understood yet and currently under
intense investigation \cite{Nakamura:00,Jeckelmann:02}. The phase
diagram in the plane $V$-$U$ is shown in Fig.~\ref{Fig1-2} (left
panel). The phase transition is of the second order up to $V
\approx 2.5$; it becomes of the first order for higher values of
$V$. The kind of phase transition that we are here analyzing is
completely different from the one usually reported in the
literature in proximity of the $U=2V$ line as this latter occurs
between an inhomogeneous charge ordered phase and an homogenous
spin ordered phase.

In Fig.~\ref{Fig1-2} (right panel) we report the phase diagram in
the plane $T$-$V$ for $U=2$, $4$ and $8$. For $U=2$ the transition
is second order for all values of $T$. For $U=4$ we can see a
first signature of a reentrant behavior. By this latter we mean a
situation in which by increasing the temperature we can first
enter and then exit a phase when within another. For $U=8$ the
reentrant behavior is clearly evident. It is interesting to
observe that the transition is continuous if there is no reentrant
behavior. When instead there is a reentrant behavior, the
transition is discontinuous up to the turning point, then becomes
continuous. It is worth noticing that the transition is clearly
marked by a discontinuity in the nearest-neighbor density-density
correlation function.

\subsection{Two-dimensional system: COM2}

\begin{figure}[tbp]
\begin{center}
\includegraphics[width=6.5cm,clip=]{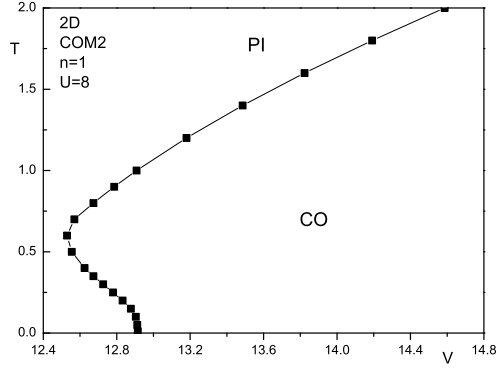}
\end{center}
\caption{The phase diagram $T$-$V$ at $U=8$.} \label{Fig3}
\end{figure}

COM2 solution for the two-dimensional case has similar
characteristics to that observed in the 1D case. The phase diagram
in the plane $T$-$V$ is reported in Fig.~\ref{Fig3}. As regards
the $V$-$U$ phase diagram, at zero temperature, we just have a
transition line with slope $\approx 3$ and the transition is
continuous for $U \le 1.8$ and first order for higher values of
$U$. For finite temperature and $U=8$ a reentrant behavior as
function of temperature is clearly observed. The transition is
first order up to the turning point $T=0.6$, then becomes
continuous. For $U \le 1.8$ no reentrant behavior is observed. The
fact that charge ordering may disappear by decreasing temperature
has been experimentally observed in $Pr
_{0.65}(Ca_{0.7}Sr_{0.3})_{0.35}MnO_3$ \cite{Tomioka:97} and
$La_{2-2x}Sr_{1+2x}Mn_2O_7$ $(0.47\le x\le 0.62)$
\cite{Chatterji:00,Dho:01}.

\subsection{Two-dimensional system: COM1}

\begin{figure}[tbp]
\begin{center}
\includegraphics[width=6.5cm,clip=]{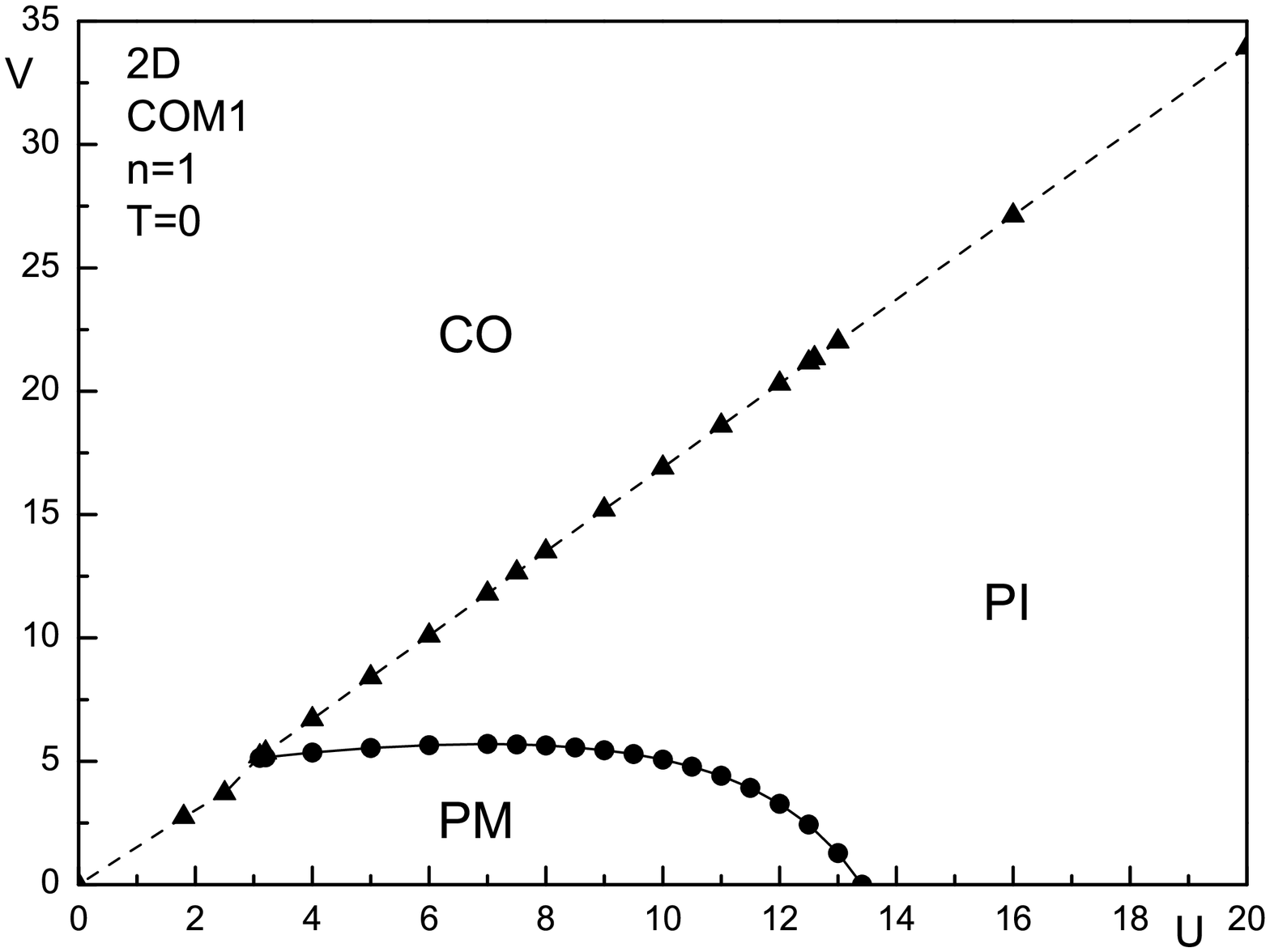}
\hfill
\includegraphics[width=6.5cm,clip=]{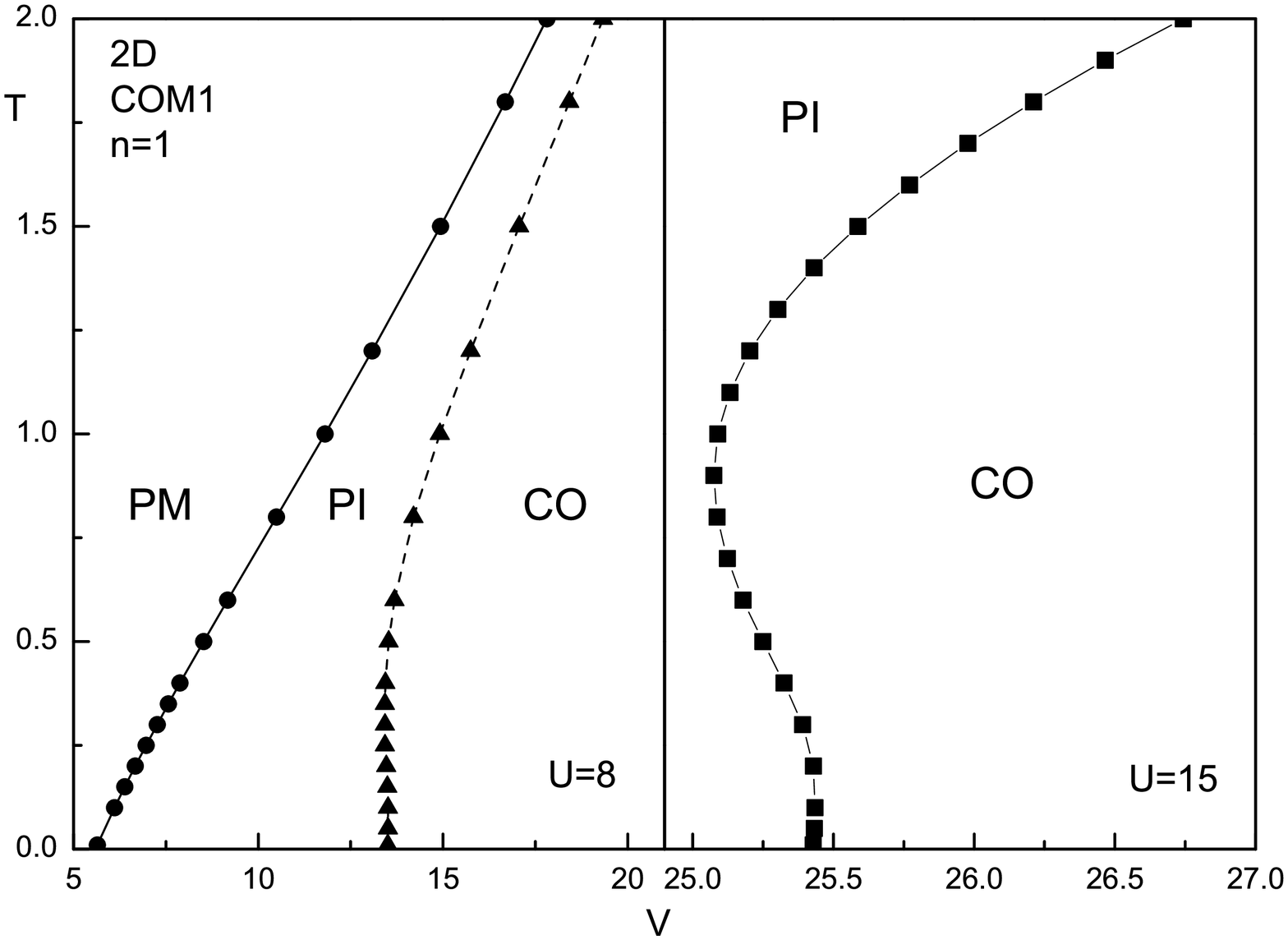}
\end{center}
\caption{(left) The phase diagram $V$-$U$ at zero temperature (PM
= paramagnetic metal). (right) The phase diagram $T$-$V$ at $U=8$
and $15$.} \label{Fig4-5}
\end{figure}

The system undergoes a metal-insulator transition at some critical
value of the onsite Coulomb potential $U$, which depends on the
intensity of the intersite potential $V$. With respect to the
$V=0$ case (the simple Hubbard model), the metallic region is
compressed by the presence of the intersite interaction and
disappears for $V>5.7$. By further increasing the intersite
Coulomb potential, the system undergoes a transition to a charge
ordered state. The complete phase diagram in the plane $V$-$U$ is
shown in Fig.~\ref{Fig4-5} (left panel). The diagram is
characterized by two critical curves, which separate the different
phases. The lower one controls the MIT, the upper one controls the
transition to a charge ordered state. The first transition is
first order for $U \le 12$ and second order for higher values of
$U$; the second one is first order for $U \ge 1.9$ and second
order for lower value of $U$.

The phase diagram in the plane $T$-$V$ is shown in
Fig.~\ref{Fig4-5} (right panel). For $U=15$ there is no metallic
phase and we have a critical temperature where a transition from
an insulating to a charge ordered state is observed. The
transition is first order up to $T=0.95$, then becomes continuous.
A reentrant behavior is observed with the same characteristics
previously discussed. For $U=8$ we have two critical temperatures,
which characterize the MIT transition and the insulator-charge
order transition, respectively. Also in this case a reentrant
behavior is observed in the latter transition.

\section{Conclusions}

The phase diagrams, in the planes $V$-$U$ and $T$-$V$, of the
half-filled extended Hubbard model, in one and two dimensions, has
been studied by means of the 2-pole approximation within the
Composite Operator Method. Transitions between the paramagnetic
(metal and insulator) phase and a charge ordered state of
checkerboard type have been found. Reentrant temperature behavior
in the plane $T$-$V$ has been observed with characteristic similar
to that experimentally found for some manganites. The rank of the
phase transitions has been studied and identified.


\begin{thebibliography}{10}
\expandafter\ifx\csname url\endcsname\relax
  \def\url#1{\texttt{#1}}\fi
\expandafter\ifx\csname urlprefix\endcsname\relax\def\urlprefix{URL }\fi

\bibitem{Emery:79}
V.~Emery, in: J.~Devreese, R.~Evrand, V.~{v}an Doren (Eds.), Highly Conducting
  One-Dimensional Solids, Plenum Press, New York, 1979, p. 247.

\bibitem{Varma:87}
C.~Varma, Sol.~Stat.~Comm. 62 (1987) 681.

\bibitem{Janner:95}
A.~Janner, Phys.~Rev.~B 52 (1995) 17158.

\bibitem{McKenzie:01}
R.~H. McKenzie, et~al., Phys.~Rev.~B 64 (2001) 085109.

\bibitem{Calandra:02}
M.~Calandra, J.~Merino, R.~H. McKenzie, Metal-insulator transition and charge
  ordering in the extended hubbard model at one-quarter filling, Phys.~Rev.~B
  66 (2002) 195102.

\bibitem{Mancini:00}
F.~Mancini, A.~Avella, Equation of motion method for composite field operators,
  Eur.~Phys.~J.~B 36 (2003) 37.

\bibitem{Avella:04}
A.~Avella, F.~Mancini, The hubbard model with intersite interaction within the
  composite operator method, to be published in Eur.~Phys.~J.~B (2004).

\bibitem{Mancini:95b}
F.~Mancini, S.~Marra, H.~Matsumoto, Spin magnetic susceptibility in the
  two-dimensional hubbard model, Physica~C 252 (1995) 361.

\bibitem{Avella:00}
A.~Avella, F.~Mancini, M.~S{\`a}nchez-Lopez, The 1d hubbard model within the
  composite operator method;, Eur.~Phys.~J.~B 29 (2002) 399.

\bibitem{Hirsch:84a}
J.~Hirsch, Phys.~Rev.~Lett. 53 (1984) 2327.

\bibitem{Nakamura:00}
M.~Nakamura, Phys.~Rev.~B 61 (2000) 16377.

\bibitem{Jeckelmann:02}
E.~Jeckelmann, Ground-state phase diagram of a half-filled one-dimensional
  extended hubbard model, Phys.~Rev.~Lett. 89 (2002) 236401.

\bibitem{Tomioka:97}
Y.~Tomioka, et~al., J.~Phys.~Soc.~Jpn. 66 (1997) 302.

\bibitem{Chatterji:00}
T.~Chatterji, et~al., Phys.~Rev.~B 61 (2000) 570.

\bibitem{Dho:01}
J.~Dho, et~al., J. Phys.: Cond. Matt. 13 (2001) 3655.

\end{thebibliography}

\end{document}